# A PROVISIONAL STUDY OF ADS WITHIN TURKIC ACCELERATOR COMPLEX PROJECT

M. Arik[1], P. S. Bilgin[2], A. Caliskan[2], M. A. Cetiner[3], S. Sultansoy[2,4*]

[1]Bogazici University, Istanbul, Turkey
[2]TOBB University of Economics and Technology, Ankara, Turkey
[3]Kastamonu University, Kastamonu, Turkey
[4]Institute of Physics of ANAS, Baku, Azerbaijan

**Abstract**

The Turkic Accelerator Complex (TAC) project has been developed with the support of the Turkish State Planning Organization by the collaboration of 10 Turkish universities. The complex is planned to have four main facilities, namely: SASE FEL Facility Based on 1 GeV Electron Linac, Third Generation Synchrotron Radiation Facility (SR) Based on 3.56 GeV Positron Synchrotron, Super-Charm Factory ($\sqrt{s}$ = 3.77 GeV) by colliding the electron beam coming from the linac with an energy of 1 GeV and positron beam coming through the positron ring with an energy of 3.56 GeV, GeV scale proton accelerator. Later has two-fold goal: Neutron Spallation Source (NSS) and ADS.
The proton accelerator construction will have 3 MeV, 100 MeV, and 1 GeV phases. The technical design report is planned to be finished in 2013. Since Turkey has essential Thorium reserves the ADS becomes very attractive for our country as emerging energy technology.

## 1. Introduction

Particle accelerators technology is one of the generic technologies which is locomotive of the development in almost all fields of science and technology. According to the U.S. Department of Energy: *Accelerators underpin every activity of the Office of Science and, increasingly, of the entire scientific enterprise. From biology to medicine, from materials to metallurgy, from elementary particles to the cosmos, accelerators provide the microscopic information that forms the basis for scientific understanding and applications* [1] (see, also, [2, 3]). For this reason, accelerator technology should become widespread all over the world. Existing situation shows that a large portion of the world, namely the South and Mid-East, is poor on the accelerator technology. UNESCO has recognized this deficit and started SESAME project in Mid-East, namely Jordan. Turkic Accelerator Complex (TAC) project is more comprehensive and ambitious project, from the point of view of it includes light sources, particle physics experiments and proton and secondary beam applications.

One of the most important applications of accelerator technology is emerging accelerator-driven systems, in other words "green" nuclear technology, which could provide the solution for the mankind's energy problem without destroying our environment [4, 5]. In this respect, proton accelerator part of the TAC project is of particular importance for Turkey.

## 2. The TAC project

Almost 20 years ago, linac-ring type charm-tau factory with synchrotron light source was proposed as a regional project for elementary particle physics [6]. Starting from 1997, a small group from Ankara and Gazi Universities begins a feasibility study for the possible accelerator complex in Turkey with the support of Turkish State Planning Organization (DPT). In 2002-2005 the conceptual design study of the TAC project was performed with a relatively enlarged group (again with the DPT support). Since 2006 TAC project activities are continued in the framework of the DPT project with participation of 10 Turkish universities.

Today, the TAC project includes:
- Linac-ring type super-charm factory
- Synchrotron light source based on positron ring
- Free electron laser based on electron linac
- GeV scale proton accelerator
- TAC test facility, namely, 20-40 MeV linac based IR FEL,

[*] E-mail of Corresponding Author: ssultansoy@etu.edu.tr

The project completion is scheduled to early 2020's (for details, see project web page: http://thm.ankara.edu.tr).

**3. The TAC proton accelerator**

At initial stage (mid 1990-ies) the proton accelerator part of the TAC project was inspired by J-PARC project (see facility web page: http://j-parc.jp ). Now, keeping in mind ADS applications, we are concentrated on GeV energy high intensity (>1mA) proton linac. Therefore, 3-5 GeV synchrotron stage is out of date. Schematic view of the TAC proton linac is shown in figure 1.

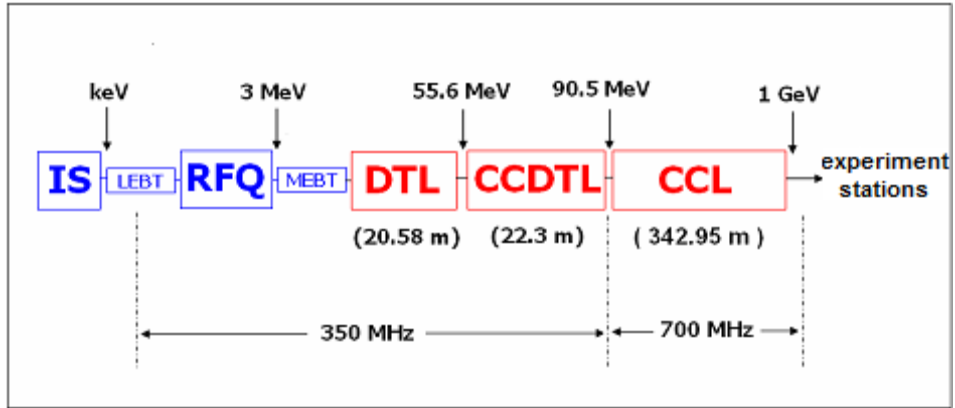

**Figure 1**. Block diagram of the TAC proton accelerator

Low and medium energy part of proposed proton accelerator will consist of normal conducting cavity structures. After an ion source, RFQ (Radio-Frequency Quadrupole), DTL (Drift Tüp Linac) and CCDTL (Coupled-Cavity Drift Tüp Linac) accelerator structures will be used, respectively. There is a low energy beam transport channel (LEBT) between the ion source and the RFQ. Its task is to match the ion beam to the RFQ by using the solenoid and steering magnets. While the beam obtained from the ion source is in the un-bunched structure, the conversion of the ion beam to the bunched structure is realized in the RFQ structure. The acceleration, the focusing and the bunching of the ion beam are simultaneously performed in the RFQ. Only transverse electric fields are used in the RFQ for these three processes. Axial longitudinal electric fields needed for the beam acceleration process are produced by internal surface modulation of the each electrode inside the RFQ. There is no acceleration in the first section of the RFQ structure called the radial matching section. After the beam is converted into the bunched structure, acceleration to energy of 3 MeV starts. The second transport section of the proton accelerator is between the RFQ and the DTL. In this channel, called medium energy beam transport (MEBT), the matching to the DTL is performed by using the quadrupole magnets. In the both transport channels, beam diagnostic elements needed such as Faraday Cup are used.

For the high energy part, using the normal conducting CCL (Coupled-Cavity Linac) cavities were planned. Beam simulations over whole line have been performed and it is shown that beam will be stable up to 60 mA beam current. For illustration, we present results for DTL in Fig. 2 [7]. It is seen that even for 50 mA emittance growth factor is less than 1.5. The normalized emittance growth along the whole linac line for the 30 mA beam current is shown in Fig. 3 [8]. Beam simulations for superconducting version of the high energy part are under development.

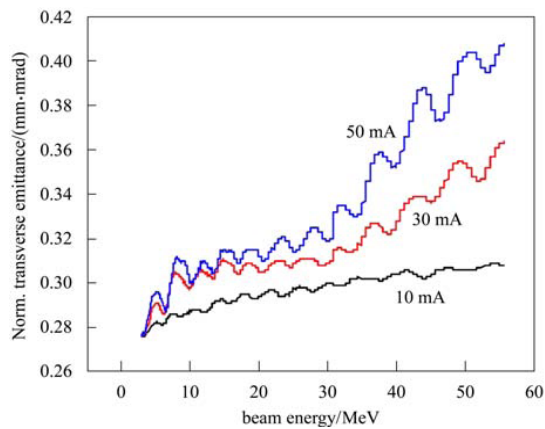

**Figure 2**. Emittance growth along the DTL for the beam currents of 10 mA, 30 mA and 50 mA, respectively.

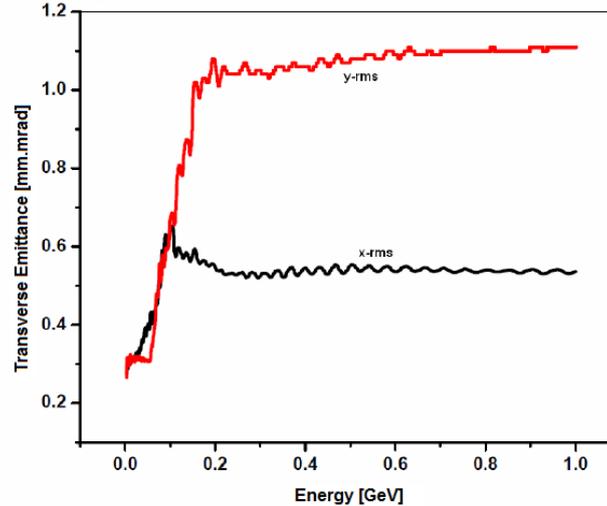

**Figure 3**. The normalized emittance growth along the whole linac line for the 30 mA beam current.

The study of TAC PA is continuing and completion of the TDR is scheduled to 2013. Alongside with ADS application, the usage of TAC PA for SNS (Spallation Neutron Source) and muon source is planned.

**4. ADS and Thorium**

It is known, that ADS has two-fold aim: burn-off of long-lived actinide waste from traditional nuclear reactors and energy production from Thorium fuel. The first one is critical for developed countries (owners of nuclear power stations, see [9] for comparison of fast reactors and accelerator driven systems for transuranic transmutation). The second one is more important, because it could provide global solution for the energy problem, especially for Third World countries.

As mentioned above, proton accelerator is crucial element of the ADS. One needs GeV energy and high intensity proton beam. While there is no problem to handle these goals separately (we have TeV energy proton beams with low intensity and high intensity proton beams with low energy), combination requires intensive R&D efforts (although, there are no visible stoppers). In Table 1 [10], we present advantages and disadvantages of existing accelerator types from the viewpoint of the ADS application.

**Table 1.** The comparison of accelerator technologies for ADS

| Technology | Cyclotron | Synchrotron | FFAG | Linac |
|---|---|---|---|---|
| Advantages | High current | High energy | High current and high energy | High current and high energy |
| Disadvantages | Energy limited | Current limited | Not yet proven | Expense |
| Examples | PSI | CERN PSB | EMMA | ESS, SNS |

Today, the favorite candidate is a proton linac. The main disadvantage of the linac option, namely, expense shall be removed at the mass production stage. The second favorite, namely, FFAG (Fixed Field Alternating Gradient) technology needs a lot of R&D efforts.

Concerning Thorium reserves in Turkey, widely shown amount is 380.000 tons [11]. However, according to [12], Turkey has in total 880.000 tons reserves (second after the Brazil, 20% of the world reserves). The last seems more realistic, keeping in mind new reserves found by AMR Corporation in Isparta region (see Corporation web page: http://amrmineralmetal.com).

**5. Conclusions**

It is well-known that Thorium is much more abundant in the nature than Uranium. Keeping in mind that U-235 isotope is used for energy production purposes; Th-232 has hundreds times more capacity in this respect. Today, there are three promising methods for Thorium utilization: molten-salt reactors [13], hybrid fusion-fission reactors (see, [14] and references therein) and ADS. The last one seems to be the most appropriate candidate for solution of the energy problem of the Mankind in foreseen future. The very important feature of the ADS is subcritical core, which principally exclude Chernobyl-like disasters.

Recently, a number of developed countries, as well as countries with essential reserves of Thorium, such as Belgium, UK, China, USA, Japan, India, Brazil, Russia and South Korea, intensified their ADS technology-related R&D activities with the aim to complete prototype systems in the near future (see, for example, [10, 15-17]). Therefore, we can expect the start of massive energy production from ADS in 2020'ies.

In the framework of the TAC project we are working on the design of proton linac for medium scale (250 MW) power generation, which corresponds to 1 GeV energy and 2.5 mA current proton beams. In our opinion, lower current will simplify reliability and availability aspects. Keeping in mind its essential Thorium reserves, Turkey should prepare corresponding state program and join to emerging international collaborations, as soon as possible.

**Acknowledgements**

We are grateful to Professor C. Rubbia for initiation of ADS research in our country almost 15 year ago and for his permanent support to Turkish scientists. We are also grateful to Professors S. Şahin and H.M. Şahin for useful discussions and organization of the special session in memoriam of Professor Engin Arık. Finally, we are grateful to Turkish State Planning Organization (DPT) and Scientific and Technological Research Council of Turkey (TUBITAK) for support.